# Indentation load-depth relation for an elastic layer with surface tension

Shao-Heng Li, Wei-Ke Yuan, Yue Ding and Gang-Feng Wang[*]

Department of Engineering Mechanics, SVL, Xi'an Jiaotong University, Xi'an

710049, China

**Abstract:** Load-depth relation is the fundamental requisite in nanoindentation tests for thin layers, however, the effects of surface tension are seldom included. This paper concerns micro-/nano-sized indentation of a bonded elastic layer by a rigid sphere. The surface Green's function with the incorporation of surface tension is first derived by applying the Hankel integral transform, and subsequently used to formulate the governing integral equation for the axisymmetric contact problem. By using a numerical method based on Gauss-Chebyshev quadrature formula, the singular integral equation is solved efficiently. Several numerical results are presented to investigate the influences of surface tension and layer thickness on contact pressure, surface deformation and bulk stress, respectively. It is found that when the size of contact is comparable to the ratio of surface tension to elastic modulus, the contribution of surface tension to the load-depth relation becomes quite prominent. With the help of a parametric study, explicit general expressions for the indentation load-depth relation as well as the load-contact radius relation are summarized, which provide the groundwork for practical applications.

---

[*] Corresponding author. E-mail Address: wanggf@mail.xjtu.edu.cn.

# 1. Introduction

Micro-/nano-indentation tests have been widely used in measuring the mechanical properties of solids e.g. metallic material [1-3], piezoelectric material [4], soft material and biological tissues [5-7]. In these tests, the relation between the measured indentation load-depth curves and the material properties is required in advance, which is closely related to the size and geometry of indenters. In general, this relationship can be obtained by investigating the corresponding contact problem. For instance, the Hertz contact model [8] and the solution of Sneddon [9] are always adopted in the analysis of indentation tests to extract the elastic modulus of materials. However, these classical contact models are sometimes not feasible because their assumptions do not hold in some practical situations.

The finite thickness of test sample is one of the most important factors that may exert a significant influence on the indentation deformation. In the last few decades, numerous studies, including both experimental researches and theoretical analyses, are delivered to examine the effects of supporting substrate on the indentation response of thin layers. By conducting nanoindentations on polycrystalline copper films, Suresh et al [10] pointed out that the overall elastic deformation and dislocation nucleation of films are sensitive to the film thickness. Saha and Nix [11] performed nanoindentation tests on the specimens of both soft films/hard substrates and hard films/soft substrates, and studied the effects of modulus mismatch on the indentation response. Through analyzing the load-depth curves of thin polymer layers, Akhremitchev and Walker [12] found large errors induced by using half space contact model. On the other hand, extensive efforts have been devoted to developing theoretical contact models for elastic layers of finite thickness. For example, some complicated asymptotic solutions [13,14] and numerical results for particular cases [5] were obtained, which are difficult to directly apply in practice. Yang [15] formulated the simplified approximate indentation load-depth relation for extremely thin layers. Dimitriadis et al [6] reconsidered the spherical indentation of an elastic layer and summarized the explicit indentation load-depth relations for layers bonded to or resting on a rigid base, which were adopted to measure the elastic modulus of thin soft layers.

Furthermore, high surface-to-volume ratio is a distinctive feature of nano-scale structures [16-20], which implies the impact of surface energy or surface tension should be considered in characterizing micro-/nano-indentations. To account for the influence of surface energy in the framework of continuum mechanics, Gurtin and Murdoch [16, 17] proposed the well-known surface elasticity model (G-M model), of which the theoretical prediction agrees well with the atomistic simulations and experimental measurements [18]. This model has been applied to study the effects of surface energy on some contact problems at nanoscale. Early in 1978, Hajji [21] originally derived the Green's functions for the elastic half space with surface tension. Using double Fourier transform technique, He and Lim [22] delivered the Green's functions for an incompressible elastic half space with surface stress. Wang and Feng [23] addressed the influence of surface tension on the elastic field of a line force acting on a half plane. On the basis of Hajji's results, Long and Wang [24] investigated the influence of surface tension on the axisymmetric contact between a rigid sphere and an elastic half space, and pointed out that the contact pressure no longer distributes like the Hertzian prediction, but is modulated by surface tension. Gao et al [25] investigated the Boussinesq problem with both surface tension and surface elasticity, and revealed that the residual surface tension is the dominant factor that affects the contact stiffness of elastic half space. In addition, Wang and Schiavone [26, 27] derived the Green's functions for anisotropic half space with surface stress effects.

To guide the nanoindentation tests of thin layers, it is necessary to develop contact models with corrections for both surface tension and layer thickness. Based on the surface elasticity theory, Zhao and Rajapakse [28] studied the effects of surface stress on the elastic fields of a two-dimensional and a three-dimensional bonded layer under a specified external traction, respectively. Subsequently, Rungamornrat et al [29] revisited the axisymmetric case with both surface tension and surface elasticity, and also pointed out that surface tension plays a dominant role for normal frictionless contact problem. Most recently, Intarit et al [30] analyzed the axisymmetric indentations of an elastic layer with surface energy. In these works, the distributions of contact stresses and contact profiles were demonstrated for some particular cases, however, the indentation load-depth relation was absent, which is the basis for indentation tests but difficult to obtain analytically. In the present paper, we aim to generalize the explicit expressions of the indentation load-depth relation and load-contact radius relation, incorporating both surface tension and finite layer thickness.

The organization of this paper is as follows. In section 2, we first derive the surface Green's function for the bonded elastic layer with surface tension, which is subsequently used to establish the contact model of a rigid sphere pressing into an elastic layer in section 3. The governing singular integral equation is solved numerically by using Gauss-Chebyshev quadrature formula. In section 4, the influences of surface tension on the distributions of contact pressure, contact deformation and bulk stress of the finite thickness layer are discussed, and the explicit expressions of the indentation load-depth relation together with the load-contact radius relation are presented.

**2. Green's function with surface tension**

According to the theory of surface elasticity [16, 17], the surface of a solid is treated as a zero-thickness membrane ideally adhering to the underlying bulk material. The governing equations for bulk material follow the classical theory of elasticity. However, the presence of surface stress leads to non-trivial constitutive and equilibrium equations on the surface. The G-M constitutive relation [16, 17] gives the surface stress $\sigma_{\alpha\beta}^s$ as

$$\begin{aligned}\sigma_{\alpha\beta}^s &= \tau_0\delta_{\alpha\beta} + 2(\mu_s - \tau_0)\varepsilon_{\alpha\beta} + (\lambda_s + \tau_0)\varepsilon_{\gamma\gamma}\delta_{\alpha\beta} + \tau_0 u_{\alpha,\beta}, \\ \sigma_{3\beta}^s &= \tau_0 u_{3,\beta},\end{aligned} \quad (1)$$

where $\tau_0$ is the surface tension independent of surface deformation, $\mu_s$ and $\lambda_s$ are the surface Lamé constants, $\varepsilon_{\alpha\beta}$ and $u_i$ are the surface strain tensor and surface displacement vector, respectively, and $\delta_{\alpha\beta}$ denotes the Kronecker delta. Throughout the paper, Einstein's summation convention is adopted for repeated Latin indices ($i, j$=1, 2, 3) and Greek indices ($\alpha, \beta$=1, 2).

Denote the unit vector normal to the surface as $\mathbf{n}(n_1, n_2, n_3)$ and the unit tangential vectors along the $x_\alpha$ direction as $\mathbf{n}^\alpha(n_1^\alpha, n_2^\alpha, n_3^\alpha)$. The equilibrium condition on the surface is described by the generalized Young-Laplace equation [31, 32]

$$\sigma^{\mathrm{s}}_{\alpha\beta,\beta} = \sigma_{ij} n_i n_j^{\alpha} - t_{\alpha}, \quad \text{(tangential direction)},$$
$$\sigma^{\mathrm{s}}_{3\beta,\beta} = \sigma_{ij} n_i n_j - t_3, \quad \text{(normal direction)}, \tag{2}$$

where $\sigma_{ij}$ is the bulk stress tensor and $\mathbf{t}(t_1, t_2, t_3)$ is the external traction vector.

Referring to the cylindrical coordinate system, we consider a three-dimensional, horizontally infinite elastic layer of finite thickness $h$ perfectly bonded to a rigid substrate, as shown in Fig. 1. The layer is a homogeneous, isotropic and linear elastic material with elastic modulus $E$ and Poisson's ratio $\nu$. At the interface $z = h$, the displacement boundary condition is written as

$$u_r = u_z = 0. \tag{3}$$

The upper surface of the layer is loaded by a uniformly distributed pressure $P_0$ over a circular region of radius $a$. Substituting Eq. (1) into the generalized Young-Laplace equation Eq. (2), one obtains the non-conventional boundary conditions on the surface $z = 0$ as,

$$k_s \left( \frac{\partial^2 u_r}{\partial r^2} + \frac{1}{r} \frac{\partial u_r}{\partial r} - \frac{u_r}{r^2} \right) + \sigma_{zr} = 0, \quad \text{(tangential direction)},$$
$$\tau_0 \left( \frac{\mathrm{d}^2 u_z}{\mathrm{d} r^2} + \frac{1}{r} \frac{\mathrm{d} u_z}{\mathrm{d} r} \right) + \sigma_{zz} + P_0 H(a-r) = 0, \quad \text{(normal direction)}, \tag{4}$$

where $k_s = 2\mu_s + \lambda_s$ is the constant of surface elasticity, and $H$ is the Heaviside step function. For normal contact problems, the influence of surface elasticity in tangential direction is relatively inessential. In this paper, we ignore the effects of surface elasticity and take $k_s = 0$.

For this axisymmetric problem, the general solution can be obtained through the Hankel integral transform. The components of displacement and stress within the bulk layer can be expressed explicitly as

$$u_r(r,z) = \frac{1+v}{E}\int_0^\infty \xi^2 \frac{\mathrm{d}G}{\mathrm{d}z} J_1(\xi r)\mathrm{d}\xi,$$

$$u_z(r,z) = \frac{1+v}{E}\int_0^\infty \xi\left[(1-2v)\frac{\mathrm{d}^2G}{\mathrm{d}z^2} - 2(1-v)\xi^2 G\right] J_0(\xi r)\mathrm{d}\xi,$$

$$\sigma_{zz}(r,z) = \int_0^\infty \xi\left[(1-v)\frac{\mathrm{d}^3G}{\mathrm{d}z^3} - (2-v)\xi^2\frac{\mathrm{d}G}{\mathrm{d}z}\right] J_0(\xi r)\mathrm{d}\xi,$$

$$\sigma_{rz}(r,z) = \int_0^\infty \xi^2\left[v\frac{\mathrm{d}^2G}{\mathrm{d}z^2} + (1-v)\xi^2 G\right] J_1(\xi r)\mathrm{d}\xi,$$  (5)

$$\sigma_{rr}(r,z) = \int_0^\infty \left\{\xi\left[v\frac{\mathrm{d}^3G}{\mathrm{d}z^3} + (1-v)\xi^2\frac{\mathrm{d}G}{\mathrm{d}z}\right] J_0(\xi r) - \frac{\xi^2}{r}\frac{\mathrm{d}G}{\mathrm{d}z} J_1(\xi r)\right\}\mathrm{d}\xi,$$

$$\sigma_{\theta\theta}(r,z) = \int_0^\infty \left\{v\xi\left[\frac{\mathrm{d}^3G}{\mathrm{d}z^3} - \xi^2\frac{\mathrm{d}G}{\mathrm{d}z}\right] J_0(\xi r) + \frac{\xi^2}{r}\frac{\mathrm{d}G}{\mathrm{d}z} J_1(\xi r)\right\}\mathrm{d}\xi,$$

where $J_n$ is the $n$-th order Bessel function of the first kind and

$$G(\xi,z) = (F_1 + F_2 z)e^{-\xi z} + (F_3 + F_4 z)e^{\xi z}. \tag{6}$$

with $F_1$, $F_2$, $F_3$ and $F_4$ being the undetermined functions of $\xi$.

Combining Eq. (5) with the boundary conditions Eqs. (3) and (4), one obtains four linear algebraic equations in terms of $F_1$, $F_2$, $F_3$ and $F_4$:

$$\begin{bmatrix} \xi & -2v & \xi & 2v \\ \dfrac{s\xi^2}{2(1-v)} + \xi & (1-2v)\left(1+\dfrac{s\xi}{1-v}\right) & \dfrac{s\xi^2}{2(1-v)} - \xi & (1-2v)\left(1-\dfrac{s\xi}{1-v}\right) \\ -\xi & 1-h\xi & \xi e^{2h\xi} & (1+h\xi)e^{2h\xi} \\ \xi & h\xi + 2(1-2v) & \xi e^{2h\xi} & [h\xi - 2(1-2v)]e^{2h\xi} \end{bmatrix} \begin{bmatrix} F_1 \\ F_2 \\ F_3 \\ F_4 \end{bmatrix}$$

$$= \begin{bmatrix} 0 \\ -\dfrac{P_0 a J_1(\xi a)}{\xi^3} \\ 0 \\ 0 \end{bmatrix}, \tag{7}$$

where $s = 2\tau_0/E^*$ with $E^* = E/(1-v^2)$ being the composite elastic modulus of the layer.

Solving the linear equation Eq. (7), one obtains

$$F_1 = \frac{Q_1}{Q}, \quad F_2 = \frac{Q_2}{Q}, \quad F_3 = \frac{Q_3}{Q}, \quad F_4 = \frac{Q_4}{Q}, \tag{8}$$

in which

$$Q_1 = \frac{aP_0\left[\nu\Lambda_1 e^{2\xi h} + h^2\xi^2 - 2h\nu\xi + 4\nu^2 - 5\nu + 2\right]J_1(\xi a)e^{2\xi h}}{\xi^4},$$

$$Q_2 = \frac{aP_0\left(\Lambda_1 e^{2\xi h} - 2\xi h + 1\right)J_1(\xi a)e^{2\xi h}}{2\xi^3},$$

$$Q_3 = -\frac{aP_0\left[\left(h^2\xi^2 + 2h\nu\xi + 4\nu^2 - 5\nu + 2\right)e^{2\xi h} + \nu\Lambda_1\right]J_1(\xi a)}{\xi^4}, \qquad (9)$$

$$Q_4 = \frac{aP_0\left[(2\xi h + 1)e^{2\xi h} + \Lambda_1\right]J_1(a\xi)}{2\xi^3},$$

$$Q = -\left[\Lambda_1\cosh(2\xi h) + \Lambda_1 s\xi\sinh(2\xi h) + 2h(h-s)\xi^2 + \Lambda_2\right]e^{2\xi h},$$

with $\Lambda_1 = 3-4\nu$, $\Lambda_2 = 8\nu^2 - 12\nu + 5$. Substituting the determined function Eq. (6) into Eq. (5) gives the elastic solution for the layer under a uniform pressure, in which both surface tension and finite thickness are accounted for.

Taking the limit $a\to 0$ with the resultant force kept constant one ($\pi a^2 P_0 = 1$), one can derive the corresponding results for a unit concentrated force. In this case, $Q_1$, $Q_2$, $Q_3$, $Q_4$ and $Q$ in Eq. (9) should be replaced by,

$$Q_1 = \frac{\left[\nu\Lambda_1 e^{2\xi h} + h^2\xi^2 - 2h\nu\xi + 4\nu^2 - 5\nu + 2\right]e^{2\xi h}}{2\pi\xi^3},$$

$$Q_2 = \frac{\left(\Lambda_1 e^{2\xi h} - 2\xi h + 1\right)e^{2\xi h}}{4\pi\xi^2},$$

$$Q_3 = -\frac{\left[\left(h^2\xi^2 + 2h\nu\xi + 4\nu^2 - 5\nu + 2\right)e^{2\xi h} + \nu\Lambda_1\right]}{2\pi\xi^3}, \qquad (10)$$

$$Q_4 = \frac{\left[(2\xi h + 1)e^{2\xi h} + \Lambda_1\right]}{4\pi\xi^2},$$

$$Q = -\left[\Lambda_1\cosh(2\xi h) + \Lambda_1 s\xi\sinh(2\xi h) + 2h(h-s)\xi^2 + \Lambda_2\right]e^{2\xi h}.$$

Then the normal displacement and bulk stress on the surface $z = 0$ are given by

$$u_z(r) = \frac{1}{\pi E^*}\int_0^\infty \frac{\left[\Lambda_1\sinh(2\xi h) - 2\xi h\right]J_0(\xi r)}{\left[\cosh(2\xi h) + s\xi\sinh(2\xi h)\right]\Lambda_1 + 2h(h-s)\xi^2 + \Lambda_2}\,d\xi, \qquad (11)$$

$$\sigma_{zz}(r) = \frac{1}{2\pi s} \int_0^\infty \left[ \Lambda_1 \cosh(2\xi h) - s\xi \Lambda_1 e^{-2\xi h} + 2h(h-s-hs\xi)\xi^2 + (1-s\xi)\Lambda_2 \right]$$
$$\times \frac{J_0(\xi r) \mathrm{d}\xi}{\left[ \cosh(2\xi h) + s\xi \sinh(2\xi h) \right] \Lambda_1 + 2h(h-s)\xi^2 + \Lambda_2} - \frac{1}{2\pi s r}. \tag{12}$$

When the thickness $h$ of layer goes to infinity, Eqs. (11) and (12) reduce to the surface Green's function of the elastic half space [21]. Moreover, when the surface tension vanishes, Eq. (11) will recover the classical solution of the contact problem for an elastic layer without surface tension [6, 13].

## 3. Contact model with surface tension

### 3.1 Governing integral equations

Based on the above surface Green's function, we formulate the axisymmetric normal frictionless contact between a rigid sphere and an elastic layer. As shown in Fig. 2, an external load $P$ is applied through the rigid sphere of radius $R$, which gives rise to an indent depth $\delta$ and a circular contact region with radius $a$. Denote the axisymmetric pressure distribution in the contact region as $p(t)$ with $t$ being the distance to the contact center. The resultant force of this pressure should equal the external load $P$, thus we have

$$\int_0^a \int_0^{2\pi} p(t) t \mathrm{d}\theta \mathrm{d}t = 2\pi \int_0^a p(t) t \mathrm{d}t = P. \tag{13}$$

By using the solution of a point force Eq. (11), one can obtain the normal displacement on the surface generated by the contact pressure $p(t)$ as

$$w(r,0) = \frac{1}{\pi E^*} \int_0^a \int_0^{2\pi} \int_0^\infty \left[ \Lambda_1 \sinh(2\xi h) - 2h\xi \right] J_0(\xi l) p(t) t$$
$$\times \frac{\mathrm{d}\xi \mathrm{d}\theta \mathrm{d}t}{\left[ \cosh(2\xi h) + s\xi \sinh(2\xi h) \right] \Lambda_1 + 2h(h-s)\xi^2 + \Lambda_2}, \tag{14}$$

with $l^2 = r^2+t^2-2rt\cos\theta$.

Similarly, the normal bulk stress on the surface can be expressed as

$$\sigma_z(r,0) = \int_0^a \int_0^{2\pi} \left\{ \int_0^\infty \left[\Lambda_1\cosh(2\xi h) - s\xi\Lambda_1 e^{-2\xi h} + 2h(h-s-hs\xi)\xi^2 + (1-s\xi)\Lambda_2\right] \right. \\ \left. \times \frac{J_0(\xi l)\mathrm{d}\xi}{\left[\cosh(2\xi h) + s\xi\sinh(2\xi h)\right]\Lambda_1 + 2h(h-s)\xi^2 + \Lambda_2} - \frac{1}{l} \right\} \frac{p(t)t}{2\pi s}\mathrm{d}\theta\mathrm{d}t. \quad (15)$$

In addition, under the assumption of small deformation, the surface normal displacement within the contact region can be described by

$$w(r,0) = \delta - \left(R - \sqrt{R^2 - r^2}\right) \approx \delta - \frac{r^2}{2R}. \quad (16)$$

Combining Eqs. (14) and (16), one can obtain the contact pressure satisfying the integral equation as

$$\frac{1}{\pi E^*} \int_0^a \int_0^{2\pi} \int_0^\infty \frac{\left[\Lambda_1 \sinh(2\xi h) - 2h\xi\right] J_0(\xi l) p(t) t \mathrm{d}\xi \mathrm{d}\theta \mathrm{d}t}{\left[\cosh(2\xi h) + s\xi\sinh(2\xi h)\right]\Lambda_1 + 2h(h-s)\xi^2 + \Lambda_2} = \delta - \frac{r^2}{2R}. \quad (17)$$

Consequently, the contact problem is completely described by the governing integral equation Eq. (17) as well as its constraint condition Eq. (13). Once the integral equations are solved, the elastic field of the bonded layer can be readily obtained by integrating the Green's function over the contact region.

### 3.2 Numerical method

Due to the complicated singular kernel in Eq. (17), it is almost impossible to derive its analytical closed-form solution, therefore numerical method is employed to solve the integral equations. Introducing a variable $\alpha = \xi l$, and then differentiating both sides of Eq. (17) with respect to $r$, one has

$$\int_0^a p(t)t \int_0^\pi \frac{r - t\cos\theta}{l} \int_0^\infty \frac{A}{B}\left(\frac{A'}{A} - \frac{B'}{B}\right) J_0(\alpha) \mathrm{d}\alpha \mathrm{d}\theta \mathrm{d}t = \frac{\pi E^* r}{2R}, \text{ for } r \leq a, \quad (18)$$

where

$$A = \sinh\left(\frac{2h\alpha}{l}\right) - \frac{2h\alpha}{\Lambda_1 l},$$

$$A' = \frac{2h\alpha}{l^2}\left[\cosh\left(\frac{2h\alpha}{l}\right) - \frac{1}{\Lambda_1}\right],$$

$$B = l\cosh\left(\frac{2h\alpha}{l}\right) + s\alpha\sinh\left(\frac{2h\alpha}{l}\right) + \frac{2h(h-s)\alpha^2}{\Lambda_1 l} + \frac{l\Lambda_2}{\Lambda_1},$$

$$B' = \left(\frac{2hs\alpha^2}{l^2} - 1\right)\cosh\left(\frac{2h\alpha}{l}\right) + \frac{2h\alpha}{l}\sinh\left(\frac{2h\alpha}{l}\right) + \frac{2h(h-s)\alpha^2}{\Lambda_1 l^2} - \frac{\Lambda_2}{\Lambda_1}.$$

(19)

Then a novel numerical method developed by Erdogan and Gupta [33] is utilized to solve the integral equation (18) with constraint condition Eq. (13). These two equations are normalized to the standard form of Gauss-Chebyshev quadrature formula by adopting the following variable transformations: $t' = 2t/a-1$, $r' = 2r/a-1$, $h' = 2h/a$, $s' = 2s/a$. Accordingly, Eqs. (18) and (13) can be rewritten as

$$\frac{1}{\pi}\int_{-1}^{1} p\left[\frac{a}{2}(1+t')\right]\sqrt{1-t'^2}K(r',t')\frac{dt'}{\sqrt{1-t'^2}} = \frac{E^*a}{4R}(1+r'), \quad -1 \leq r' \leq 1, \quad (20)$$

$$\frac{1}{\pi}\int_{-1}^{1} p\left[\frac{a}{2}(1+t')\right]\sqrt{1-t'^2}(1+t')\frac{dt'}{\sqrt{1-t'^2}} = \frac{2P}{\pi^2 a^2}, \quad (21)$$

where

$$K(r',t') = (1+t')\int_0^\pi \frac{(1+r')-(1+t')\cos\theta}{l'}\int_0^\infty \frac{A}{B}\left(\frac{A'}{A} - \frac{B'}{B}\right)J_0(\alpha)d\alpha d\theta, \quad (22)$$

with

$$A = \sinh\left(\frac{2h'\alpha}{l'}\right) - \frac{2h'\alpha}{\Lambda_1 l'},$$

$$A' = \frac{2h'\alpha}{l'^2}\left[\cosh\left(\frac{2h'\alpha}{l'}\right) - \frac{1}{\Lambda_1}\right],$$

$$B = l'\cosh\left(\frac{2h'\alpha}{l'}\right) + s'\alpha\sinh\left(\frac{2h'\alpha}{l'}\right) + \frac{2h'(h'-s')\alpha^2}{\Lambda_1 l'} + \frac{l'\Lambda_2}{\Lambda_1}, \quad (23)$$

$$B' = \left(\frac{2h's'\alpha^2}{l'^2} - 1\right)\cosh\left(\frac{2h'\alpha}{l'}\right) + \frac{2h'\alpha}{l'}\sinh\left(\frac{2h'\alpha}{l'}\right) + \frac{2h'(h'-s')\alpha^2}{\Lambda_1 l'^2} - \frac{\Lambda_2}{\Lambda_1},$$

$$l' = \sqrt{(1+r')^2 + (1+t')^2 - 2(1+r')(1+t')\cos\theta}.$$

According to Gauss-Chebyshev quadrature formula, Eqs. (20) and (21) can be further transformed into linear algebraic equations

$$\mathbf{R}_{n\times n}\mathbf{Y}_{n\times 1} = \mathbf{f}_{n\times 1}, \quad (24)$$

where

$$R_{ij} = \frac{1}{n}K(r_i', t_j'), \quad R_{nj} = \frac{1}{n}(1+t_j'), \quad (25)$$

$$Y_j = p\left[\frac{a}{2}(1+t_j')\right]\sqrt{1-t_j'^2}, \quad (26)$$

$$f_i = \frac{E^*a}{4R}(1+r_i'), \quad f_n = \frac{2P}{\pi^2 a^2}, \quad (27)$$

$$r_i' = \cos\left(\frac{i\pi}{n}\right), \quad t_j' = \cos\left[\frac{(2j-1)\pi}{2n}\right], \quad 1 \le i \le n-1, \quad 1 \le j \le n. \quad (28)$$

For a specific indentation situation under a given load $P$, the contact radius $a$ can be determined through an iterative approach [24]. The Hertzian radius $a_H = [3PR/(4E^*)]^{1/3}$ is used as the initial trial value, then the corresponding pressure distribution can be obtained by solving Eq. (24). The next trial value should be chosen according to the form of the obtained pressure distribution. If the pressure tends to be negative at contact fringe, then this trial value of contact radius exceeds the real one and a smaller one should be chosen at the next step. Otherwise, if the fringe pressure goes to infinity, then this trial value is much smaller than the real one, and we choose a larger one at next step. The final obtained pressure should be continuous within the whole contact region, which corresponds to the real value of contact radius. Performing bisection method continuously, one can calculate the contact radius and contact pressure to a high accuracy.

## 4. Results and Discussion

By applying the numerical method in section 3.2, we succeed in solving the contact problem. From the dimensionless form of the governing integral equations, it is found that the influence of surface tension and layer thickness can be indicated by the values of $a/s$ and $a/h$, respectively. Different from that of elastic half space, the solution of indentation of finite thickness layer changes with the Poisson's ratio. In this paper, we present the results for $\nu = 0.4$ as an example.

For different values of $a/s$, the contact pressure $p(t)$ normalized by average pressure is plotted in Fig. 3(a), with $a/h$ kept constant one. It can be seen that the contact pressure tends to be uniformly distributed as $a/s$ decreases, and reduces to the classical results when $a/s$ is sufficiently large. This variation of contact pressure with respect to $a/s$ is qualitatively consistent with that of half space [24]. On the other hand, to investigate the impact of layer thickness on the distribution of contact pressure, results for different values of $a/h$ are plotted in Fig. 3(b), with $a/s$ kept constant one. When $a/h$ reaches a vanishing value, one can find that the contact pressure approaches the solution of Long et al [24] for indentation on a half space. However, when the value of $a/h$ increases, the pressure distribution will evidently deviate from that of half space, and it is necessary to consider the influence of layer thickness.

Fig. 4(a) and 4(b) display the distributions of normalized displacement on the surface for various values of $a/s$ (with $a/h = 1$) and $a/h$ (with $a/s = 1$), respectively. As can be expected, the normal displacement distribution approaches to the classical results as the surface tension diminishes, and reduces to that of elastic half space case as the layer thickness rises to infinity. From Fig. 4(a), the displacement within the contact region decreases with the decline of $a/s$. On the other hand, it can be clearly found that the normalized displacement on the surface will definitely decrease as the elastic layer gets thinner, as shown in Fig. 4(b). When the contact radius is larger than the layer thickness, the influence of rigid substrate on the layer deformation will be quite evident. In addition, the variations of normalized bulk stress on the surface with respect to $a/s$ and $a/h$ are reported in Fig. 5(a) and 5(b), respectively, which shows continuous slopes across the contact edge. The existence of surface tension leads to a smoother stress distribution and the supporting substrate leads to a higher stress within contact region.

Thereby, the analysis of load-depth curves in nanoindentation tests for thin layered samples should take notice of the effects of both surface tension and layer thickness. Here, substantial indentation cases for various values of $a/s$ and $a/h$ are numerically computed, of which the results are then fitted to summarize the overall indentation response relations.

First, the relationship between external load and contact radius is carried out. As discussed above, two dimensionless variables $a/s$ and $a/h$ can be chosen to indicate the influence of surface tension and layer thickness, respectively. Fig. 6 plots the variation of normalized load with respect to $a/h$ [0.01, 2] and $a/s$ [0.2, 50]. It is found that the load-contact radius relation can be well fitted by

$$\frac{P}{P_H^{(a)}} = K_1 + K_2 \cdot \left(\frac{a}{s}\right)^{K_3}, \tag{29}$$

where $P_H^{(a)} = 4E^* a^3/R/3$, $K_1$, $K_2$ and $K_3$ are functions of $\zeta = a/h$, as depicted in Fig. 7, and they can be fitted by polynomial as,

$$K_1 = 1 + 0.128\zeta + 0.231\zeta^2, \tag{30}$$

$$K_2 = 2.743 + 0.042\zeta + 1.220\zeta^2 - 0.683\zeta^3 + 0.130\zeta^4, \tag{31}$$

$$K_3 = -0.934 - 0.022\zeta + 0.157\zeta^2 - 0.095\zeta^3 + 0.019\zeta^4. \tag{32}$$

If the layer thickness tends to be infinity i.e. $\zeta \to 0$, these functions reduce to three constants: $K_1 = 1$, $K_2 = 2.743$, $K_3 = -0.934$, which agree with the fitted relation of half space [34]. Moreover, if the contact radius $a$ is much larger than the intrinsic length of the layer material $s$, i.e. $a/s \to \infty$, Eq. (29) becomes

$$P = \frac{4E^* a^3}{3R}\left(1 + 0.128\zeta + 0.231\zeta^2\right), \tag{33}$$

which agrees well with the conventional result for elastic bonded layers [5], as shown in Fig. 8.

Next, the indentation load-depth relation is finally presented. Fig. 9 shows the variation of normalized load with respect to $(R\delta)^{0.5}/h$ and $(R\delta)^{0.5}/s$. Similarly, in terms of indent depth, the external load can be expressed as

$$\frac{P}{P_H^{(\delta)}} = L_1 + L_2 \left(\frac{\sqrt{R\delta}}{s}\right)^{-0.86}, \tag{34}$$

where $P_H^{(\delta)} = 4E^*\delta^{1/2}R^{3/2}/3$, $L_1$ and $L_2$ are functions of $\chi = (R\delta)^{0.5}/h$. As shown in Fig. 10, the functions $L_1$, $L_2$ in the range of $[0.01 \leq (R\delta)^{0.5}/h \leq 1.2, 0.13 \leq (R\delta)^{0.5}/s \leq 50]$ can be fitted by polynomial as

$$L_1 = 1 + 0.667\chi + 2.217\chi^2 - 0.504\chi^3, \tag{35}$$

$$L_2 = 0.783 + 1.82\chi - 1.067\chi^2 + 0.24\chi^3. \tag{36}$$

Again, when the thickness tends to be infinity, these polynomial functions reduce to two constants, $L_1 = 1$, $L_2 = 0.783$, which reduces to the empirical relation for indentation of an half space with surface tension [34]. Furthermore, when the effect of surface tension is ignored, Eq. (34) becomes

$$P = \frac{4}{3}E^*R^{1/2}\delta^{3/2}\left(1 + 0.667\chi + 2.217\chi^2 - 0.504\chi^3\right). \tag{37}$$

As shown in Fig. 11, this equation is also in good agreement with the classical result for bonded elastic layers [5, 6], and comes back to the Hertzian contact solution as the layer thickness goes to infinity.

## 5. Conclusions

In summary, we have investigated the indentation of a rigid sphere on a bonded elastic layer of finite thickness with surface tension. All the numerical results reveal that it is imperative to account for the effects of surface tension and finite thickness in the analysis of nanoindentation test. When the contact radius is on the same magnitude as the layer thickness, the existence of rigid substrate evidently reduces the layer deformation and leads to a higher stress within contact region. On the other hand, compared with the conventional solution, the contact pressure of an elastic layer with surface tension tends to distribute uniformly as the characteristic length of indentation declines. When the contact radius is on the same magnitude as the ratio of surface tension to elastic modulus, the existence of surface tension also highly decreases the layer deformation but leads to a smoother stress distribution. Thus, both surface tension and finite thickness make the elastic layer perform stiffer in nanoindentation tests. To be quantitative, we develop necessary corrections for the expressions between indentation load and contact radius as well as that between the load and indent depth, in which the influences of surface tension and finite thickness are included. These general relations can be employed in the determination of elastic properties of soft thin samples by nanoindentation tests.

**Acknowledgement**

This work was supported by the National Natural Science Foundation of China (grant number 11525209)

**Figure captions:**

Fig. 1. Schematic diagram of a uniform pressure acting on an elastic layer bonded to rigid substrate.

Fig. 2. Contact between a rigid sphere and an elastic layer with surface tension.

Fig. 3. Contact pressure distribution for (a) $a/h = 1$ and (b) $a/s = 1$.

Fig. 4. Normal displacement distribution on the surface for (a) $a/h = 1$ and (b) $a/s = 1$.

Fig. 5. Normal bulk stress distribution on the surface for (a) $a/h = 1$ and (b) $a/s = 1$.

Fig. 6. Variation of the normalized load $P/P_H^{(a)}$ with respect to $a/h$ and $a/s$.

Fig. 7. Fitting curves of the dimensionless functions: (a) $K_1(\zeta)$, (b) $K_2(\zeta)$ and (c) $K_3(\zeta)$.

Fig. 8. Comparison of the normalized load $P/P_H^{(a)}$ with the results of Ref. [5].

Fig. 9. Variation of the normalized load $P/P_H^{(\delta)}$ with respect to $(R\delta)^{0.5}/h$ and $(R\delta)^{0.5}/s$.

Fig. 10. Fitting curves of the dimensionless functions: (a) $L_1(\chi)$ and (b) $L_2(\chi)$.

Fig. 11. Comparison of the normalized load $P/P_{\mathrm{H}}^{(\delta)}$ with the results of Refs. [5,6].

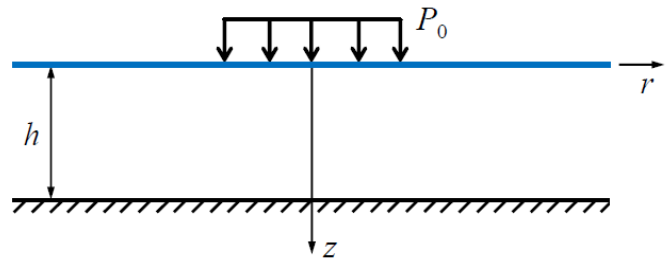

Figure 1

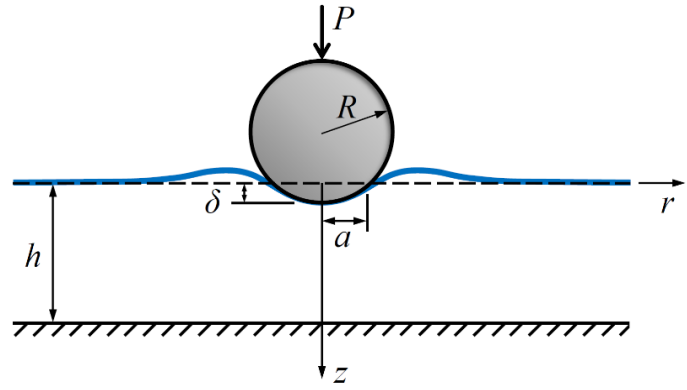

Figure 2

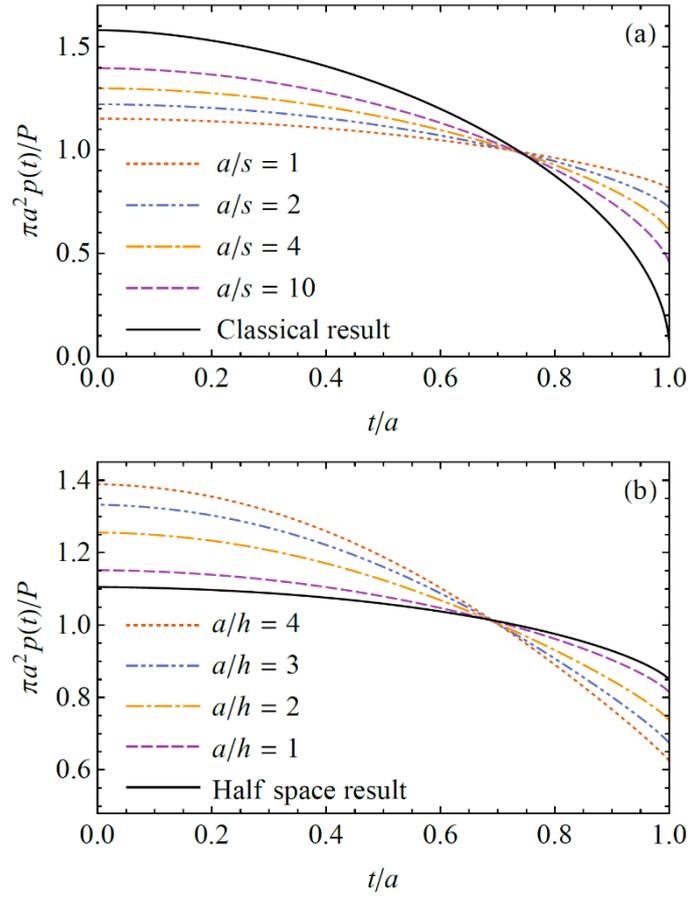

Figure 3

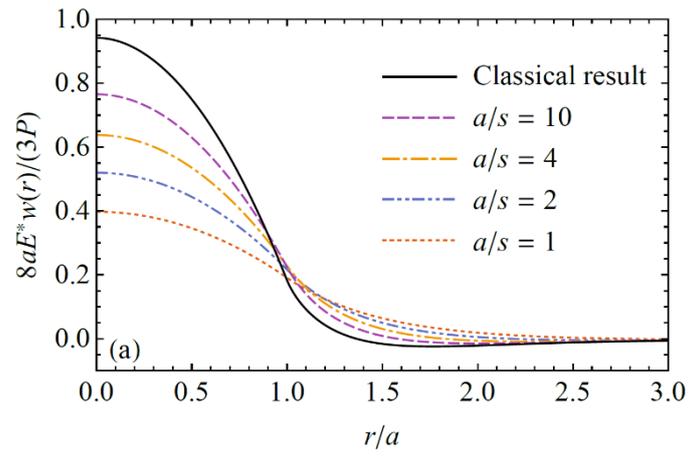

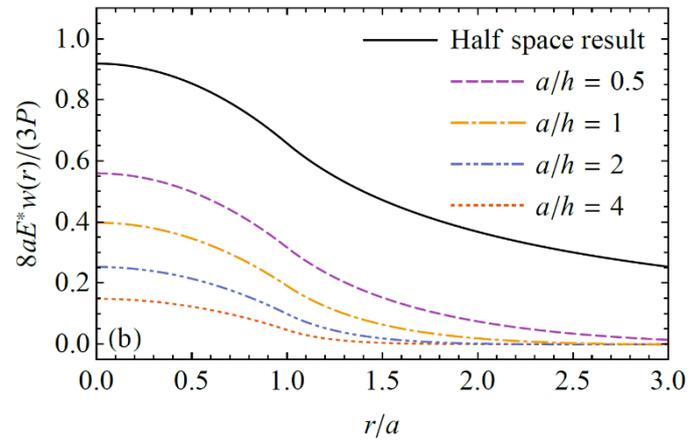

Figure 4

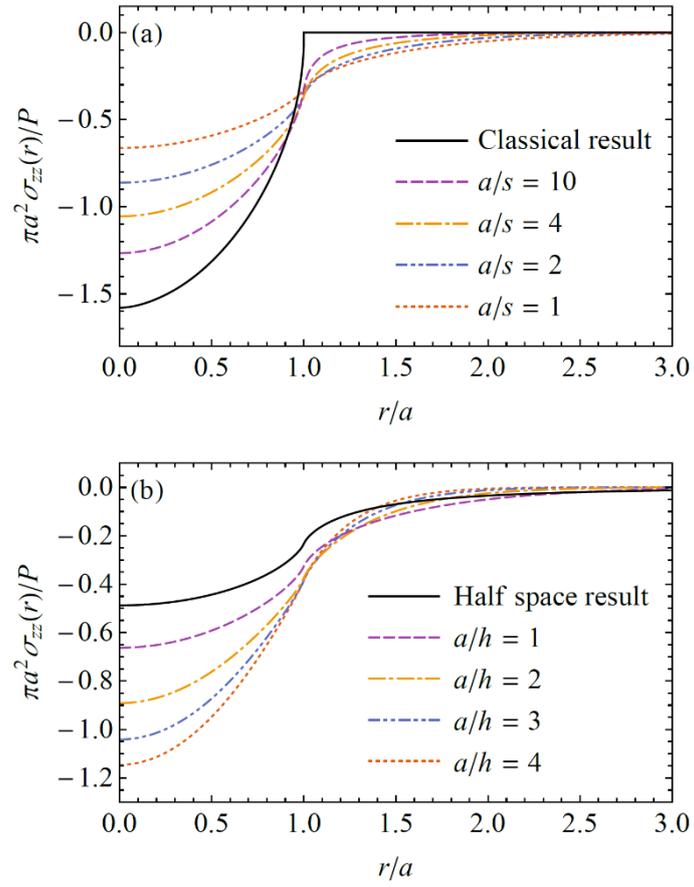

Figure 5

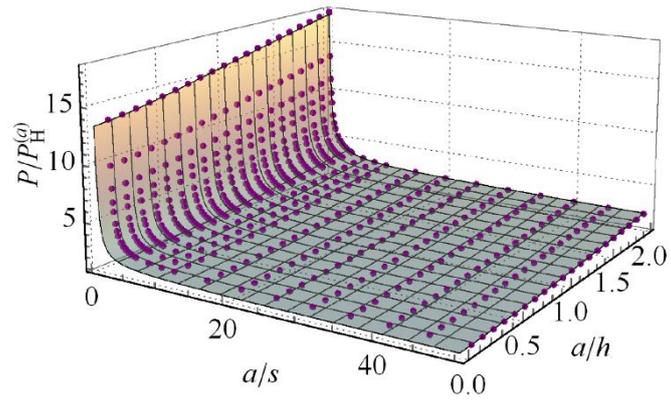

Figure 6

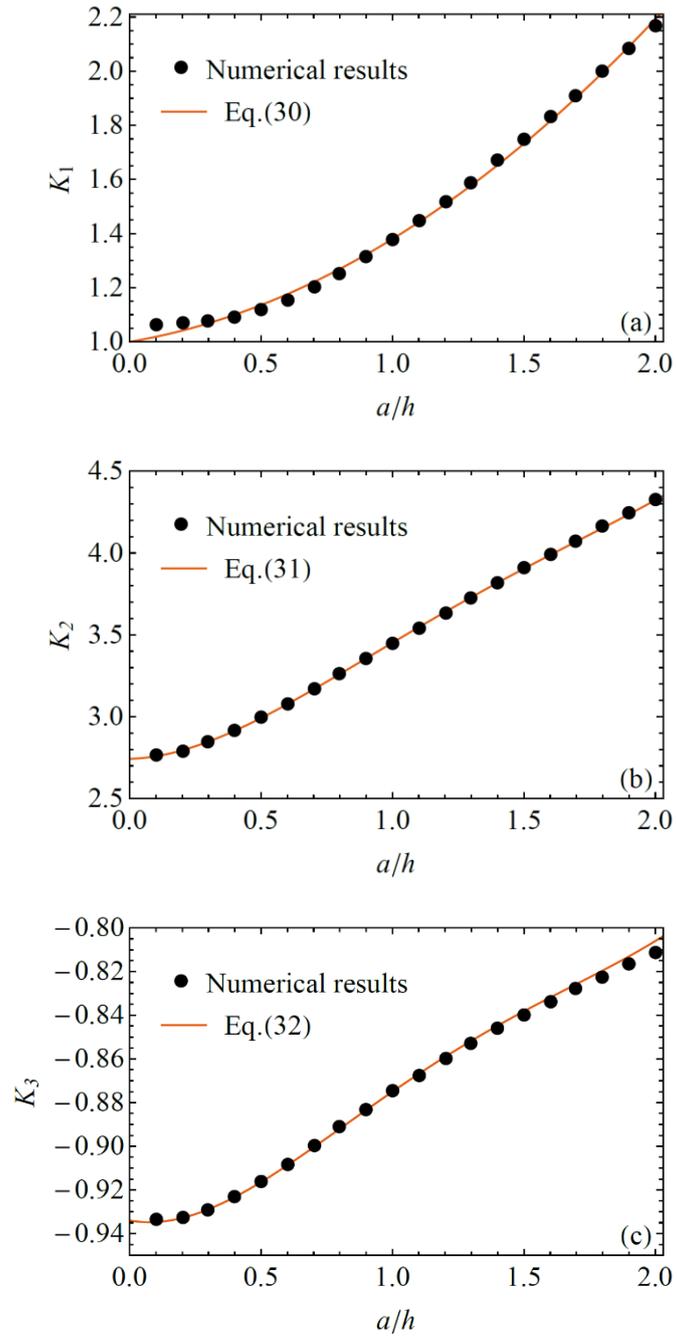

Figure 7

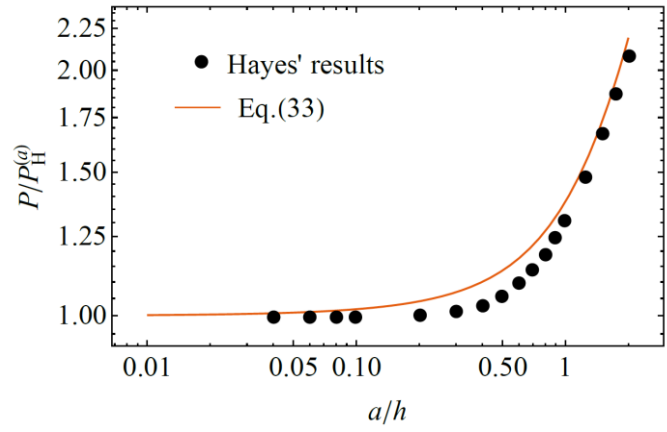

Figure 8

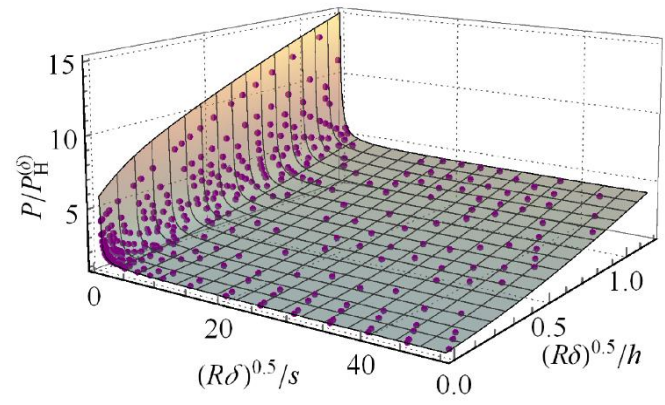

Figure 9

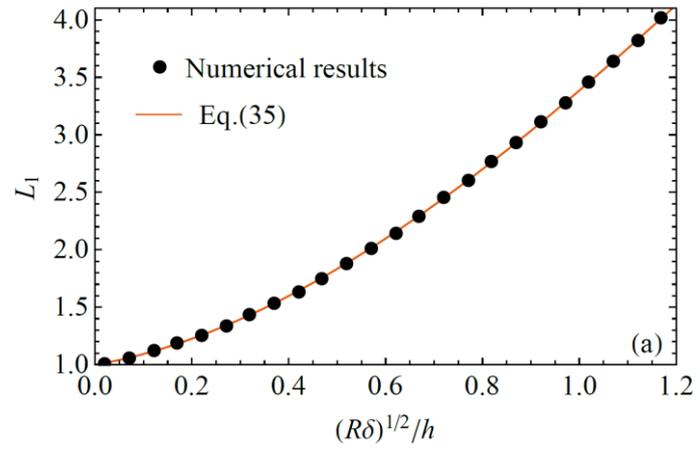

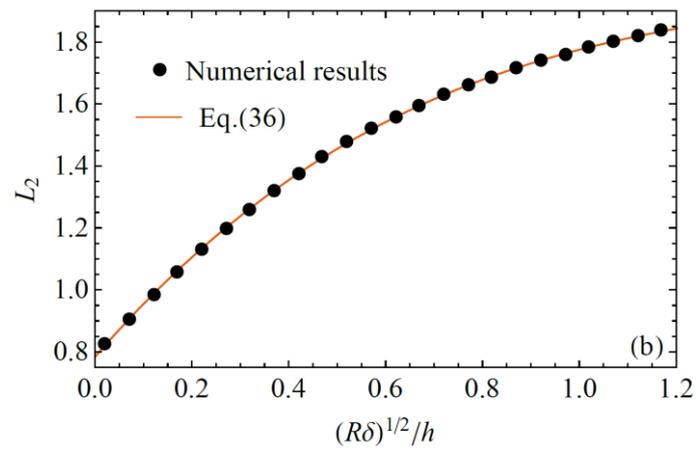

Figure 10

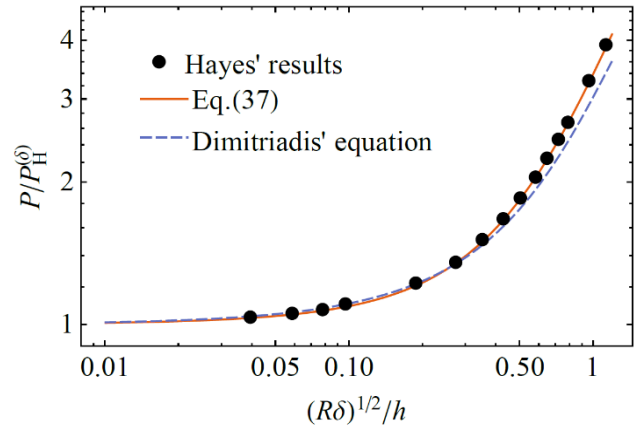

Figure 11